\documentclass[referee]{aa}     % LaTeX A&A  Standard Fonts
\usepackage{amssymb}
\usepackage{epsfig}
\usepackage{times}
\usepackage{float}
\begin{document}
\thesaurus{11.03.1,         % Galaxies: clusters: general
           12.04.2         % Cosmology: diffuse radiation
}
%\titlerunning{Thermal equilibrium}
\titlerunning{Asphericity of galaxy clusters and the SZ effect}
\title{Asphericity of galaxy clusters and the Sunyaev-Zel'dovich effect}
\author{Denis Puy$^{\dagger \ast}$, Lukas Grenacher$^{\dagger \ast}$, 
Philippe Jetzer$^{\dagger \ast}$, Monique Signore$^\ddagger$}
\authorrunning{Puy et al.}
\offprints{puy@physik.unizh.ch}
\institute{
$^\dagger$Paul Scherrer Institute\\ 
Laboratory for Astrophysics\\
CH-5232 Villigen PSI (Switzerland)
\\
\\
$^\ast$Institute of Theoretical Physics\\
University of Z\"urich\\
Winterthurerstrasse, 190\\
CH-8057 Z\"urich (Switzerland)
\\
\\
$^\ddagger$Observatoire de Paris\\
D\'epartement de Radioastronomie Millim\'etrique\\
61, Avenue de l'Observatoire\\
F-75014 Paris (France)}
\date{Received 27 July 2000 / Accepted 31 August 2000}
\maketitle
\begin{abstract}
In this paper we investigate the Sunyaev-Zel'dovich (SZ) effect 
and the X-ray surface brightness for clusters of galaxies with a 
non-spherical mass distribution. In particular, 
we consider the influence of the shape and the finite extension 
of a cluster, as well as of a polytropic thermal profile on the Compton 
parameter, the X-ray surface brightness and on the determination of the 
Hubble constant.
We find that the non-inclusion of such effects can induce errors
up to 30\% in the various parameters and in particular on the 
Hubble constant value, when compared with results obtained under
the isothermal, infinitely extended and spherical shape assumptions.

\end{abstract}
%----------------------------------------------------------------------------
\section{Introduction}
%----------------------------------------------------------------------------
Over the last few years, studies on galaxy clusters using X-ray emission 
observations have been a source of a tremendous increase in the litterature, 
especially those using Sunyaev-Zel'dovich (SZ) effect. The SZ effect (Sunyaev \& Zel'dovich \cite{Sun72}, Rephaeli \cite{Rep95}, 
Birkinshaw \cite{Bir99}) is one of the major sources of secondary anisotropies 
of the Cosmic Microwave Background (CMB) arising from inverse Compton 
scattering of the microwave photons by hot electrons in 
clusters of galaxies. 
\\
Many different works have been developed during recent years leading to 
the use of this effect for studies of cosmology (Bernstein \& Dodelson \cite{Ber90}; Aghanim, de Luca, 
Bouchet et al. \cite{Agh97}; Barbosa, Bartlett, Blanchard  et al. \cite{Bar96}; 
Bartlett, Blanchard \& Barbosa \cite{Bar98}; Cooray \cite{Coo98}). 
Observations in the millimetre and submillimetre wavebands (Perrenod \& 
Lada  \cite{Per79}; Chase et al. \cite{Cha87}; 
Silverberg et al. \cite{Sil97}) 
give important information on the characteristics 
of clusters of galaxies. For 
example, by combining the SZ intensity change and the $X$-ray emission 
observations, and solving for the number density distribution of electrons 
responsible for both these effects (after assuming a certain geometrical 
shape), the angular diameter distance to galaxy clusters can be derived. 
Assuming a cosmological model, this leads to an estimate of 
the Hubble constant $H_0$ (Holzapfel, Arnaud, Ade et al., 
\cite{Hol97} for Abell 2163; Birkinshaw \& Hughes, \cite{Bir94} for Abell 2218).
\\
The SZ effect thus offers the possibility to put important constraints on the 
cosmological models. For this reason,  different projects to 
measure the SZ effect are under way for example the MITO instrument 
(De Petris, Aquilina, Canonico et al., \cite{Dep96}) or (the longer term) the 
Planck mission (ESA report \cite{Esa97}).
\\
The SZ effect is difficult to measure accurately, since systematic errors 
can be significant. For instance, Inagaki, Suginohara \& Suto (\cite{Ina95}) 
made an analysis of the 
reliability of the Hubble constant determination based on the SZ effect. 
\\
An additional effect arises if the cluster
has a peculiar velocity (kinematic effect). Several papers 
discussed the influence of the kinematic effect on the measurement of the
thermal SZ effect (De Luca, D\'esert \& Puget, \cite{Del95}; 
Audit \& Simmons, \cite{Aud99} for transverse clusters velocities). 
Note that the kinematic effect 
can thus be used to infer the peculiar velocity of clusters of galaxies, 
if the value of the Hubble constant is known (Rephaeli \& Lahav, \cite{Rep91}; 
Haehnelt \& Tegmark, \cite{Hae96}). Another possible distortion on the SZ 
effect is due to gravitational lensing (Blain, \cite{Bla98}; 
Roettiger et al., \cite{Roe97}).
\\
The extension and the geometry of hot gas distribution in clusters of 
galaxies is also an important source of systematic errors in the SZ effect. 
Cooray (\cite{Coo98}) showed that projection effects of clusters can affect 
the calculations of the Hubble constant and the gas mass fraction. Recently, 
Sulkanen (\cite{Sul99}) showed that galaxy cluster 
shapes can produce systematic errors in $H_0$ measured via the SZ effect. 
It is thus necessary to know, at least approximately, the shape of the 
clusters, for instance, if they are oblate or prolate 
(Cooray \cite{Coo2000}, Hughes \& Birkinshaw \cite{hubi98}) or have a more general 
geometry.
\\
The $\beta$-model (Cavaliere \& Fusco-Femiano \cite{Cav76}) is 
widely used in $X$-ray astronomy to 
parametrize the gas density profile in clusters of galaxies by fitting their 
surface brightness profile. Nevertheless, fitting an aspherical distribution 
with a spherical $\beta$-model can lead to an important inaccuracy  
(see Inagaki, Suginohara \& Suto \cite{Ina95}).\\
The aim of this paper is to 
investigate the influence of the shape and the finite extension of an 
ellipsoidal cluster gas 
distribution on the SZ effect, and to discuss the possible errors 
induced in the inferred value for $H_0$. The paper is organized as follows:
\\
In Section \ref{SZ:S2} we present the calculations of the SZ effect and the $X$-ray 
surface brightness for an ellipsoidal shape with an isothermal 
profile and a finite cluster extension. Details 
of the calculations are reported in two appendices.
\\
Section \ref{SZ:S3} is then devoted to a quantitative discussion of the 
incidence of these
effects on the SZ measurements, in particular, the finite extension and the
geometry (prolate and oblate) of the cluster.
The influence of a polytropic thermal profile on the SZ measurements is 
also considered.
\\
The discussion and conclusion are given in Section \ref{SZ:S4}. 

%------------------------------------------------------------------------
\section{Basic equations of the SZ effect and the X-ray surface brightness 
for an ellipsoidal geometry}
\label{SZ:S2}
%------------------------------------------------------------------------

Different $X$-ray surface brightness measurements in 
clusters of galaxies clearly indicate an asphericity of the cluster shape. 
Fabricant, Rybicki \& 
Gorenstein (\cite{Fab84}) showed a pronounced ellipticity for the cluster 
Abell 2256 which indicates that the underlying density profile is 
aspherical. Allen et al. (\cite{All93}) reached the 
same conclusion for the profile of Abell 478, and Hughes, Gorenstein \& 
Fabricant (\cite{Hug88}) for the Coma cluster. It is thus of relevance to 
study the influence of non-spherical shape on the results of 
clusters which have been reported so far.
\\
Given the above results, we have assumed an ellipsoidal $\beta$ model:
\begin{equation}
n_e(r_x, r_y, r_z) = 
n_{eo} \left[ 1 + \frac{r_x^2}{\zeta_1^2} + \frac{r_y^2}{\zeta_2^2}
+ \frac{r_z^2}{\zeta_3^2} \right ]^{-3 \beta /2 }~,\label{eq:dp}
\end{equation}
where $n_{eo}$ is the electron number density at the center of the cluster and 
$\beta$ is a free fitting parameter, which lies in the range $1/2 
\leq \beta \leq 1$. The set of coordinates $r_x$, $r_y$ and $r_z$, as well as 
the half axes of the ellipsoid, $\zeta_1$, $\zeta_2$ and $\zeta_3$, are defined
in units of the core radius $r_c$ of the corresponding spherical shape.

The fractional temperature decrement $\Delta T_{SZ}$ of the cosmic microwave 
background due to the SZ effect is expressed as
\begin{equation}
\Delta T_{SZ} = -f(\omega)~y 
\end{equation}
with
\begin{equation}
f(\omega) = \left[ \frac{\omega (e^{\omega} + 1)}{e^{\omega}-1} 
- 4 \right], \ \ {\rm where} \ \  \omega = \frac{h \nu}{ k_B T_{CMB}} \ \ 
{\rm and} 
\end{equation}
$T_{CMB}$ is the temperature of the 
cosmic background radiation at $z=0~$ ($T_{CMB} = 
2.728 \pm 0.002$ K, Fixsen et al. \cite{Fix96}) and 
$k_B$ the Boltzmann constant. The 
Compton parameter $y$ is defined as
\begin{equation}
y  = 2  r_c \, \int_{0}^{l} \, \frac{k_B T_e}{m_e c^2} \sigma_T n_e dr_y~.
\end{equation}
We have chosen the line of sight to be along the $r_y$ axis. $l$ is the 
maximal extension of the hot gas in units of the core radius $r_c$
 along the line of sight, $T_e$ the temperature of the hot gas, $m_e$ 
the electron mass, $\sigma_T$ the Thomson cross section and $c$ the speed of 
light. 
\\
The $X$-ray surface brightness of a cluster is given by:
\begin{equation}
S_X = \frac{r_c}{2 \pi (1 + z)^3} \, \int_{0}^{l} \, n_e^2 \epsilon _X 
~dr_y~,
\end{equation}
where $z$ is the redshift of the cluster, which takes into account the 
cosmological transformation of the spectral surface brightness,
 $\epsilon _X$ the spectral emissivity of the gas, which can be approximated 
by a typical value (for $T_e\geq 3\times 10^7$ K, 
Sarazin \cite{Sar86}):
\begin{equation}
\epsilon_X = \epsilon \, \sqrt{T_e}~{\rm with}~\epsilon\approx 
3.0\times 10^{-27}~n_p^2\,\,{\rm erg~cm^{-3}~s^{-1}}~,
\end{equation}
where $n_p$ denotes the proton number density.
The thermal SZ effect and the $X$-ray surface brightness depend 
on the temperature profile and of course on the density profile. We 
consider in the following an isothermal profile. The influence of a polytropic
thermal profile is discussed in Section 3.
%---------------------------------------
\subsection{Isothermal profile}
%---------------------------------------
In this case (with $T_e = T_{eo}$) the Compton parameter and the surface 
brightness depend only on the density profile
\begin{equation}
y =  
\frac{k_BT_{eo} n_{eo}r_c \sigma_T}{m_e c^2}  \, 
I_{SZ} \ \ {\rm with} \ \ I_{SZ}
= 2 \, \int_{0}^{l} \, \frac{n_e}{n_{eo}} dr_y ~,
\end{equation}
\begin{equation}
S_X =  \frac{\epsilon \sqrt{T_{eo}} n_{eo}^2 r_c}{4 \pi 
(1 + z)^3} \, I_{S_X} \ \ {\rm with} \ \ I_{S_X}
 =  2 \, \int_{0}^{l} \, \left( \frac{n_e}{n_{eo}} 
\right)^2 \, dr_y ~,
\end{equation}
where we introduce the {\it structure integrals} $I_{SZ}$ and $I_{S_X}$, 
which depend only on the geometry and the extension $l$ of the cluster 
along the line of sight $r_y$. 
\\
With the structure integrals calculated in Appendix A, we obtain for $y$ 
and $S_X$:
\begin{eqnarray}
y(r_x,r_z) &=& 
 \frac{k_BT_{eo} \sigma_T n_{eo} \zeta_2 r_c}
{m_e c^2}  
 \times   \left( 1+\frac{r_x^2}{\zeta_1^2} + \frac{r_z^2}{\zeta_3^2} 
\right)^{-\frac{3}{2}\beta + \frac{1}{2}} \nonumber \\
&\times& 
\left[ B\left(\frac{3}{2}\beta-\frac{1}{2},\frac{1}{2}\right) -B_m\left(\frac{3}{2}\beta-
\frac{1}{2},
\frac{1}{2}\right)\right]
\label{eq:dt}
\end{eqnarray}
\begin{eqnarray}
S_X(r_x,r_z)&=& 
\frac{\epsilon n_{eo}^2 \sqrt{T_{eo}} \zeta_2 r_c}{4 \pi 
(1 + z)^3}\times 
\left( 1+\frac{r_x^2}{\zeta_1^2} + \frac{r_z^2}{\zeta_3^2} 
\right)^{-3\beta + \frac{1}{2}} \nonumber \\
&\times& 
\, \left[ B\left(3\beta-\frac{1}{2},\frac{1}{2}\right) 
-B_m\left(3\beta-\frac{1}{2},
\frac{1}{2}\right)\right]~,
\label{eq:x}
\end{eqnarray}
where the cut-off parameter $m$, which depends also on the extension $l$ of 
the cluster, is:
\begin{equation}
m =  \frac{1+(r_x/\zeta_1)^2 + (r_z/\zeta_3)^2}
{1+(r_x/\zeta_1)^2 + (r_z/\zeta_3)^2 + (l/\zeta_2)^2}~.
\end{equation}
For the ratio between $y^2$ 
and $S_X (r_x,r_z)$ we thus obtain
\begin{eqnarray}
\frac{y^2(r_x, r_z)}{S_X (r_x,r_z)} \,& =& \, 
\lambda ~ T_{eo}^{3/2} \, \zeta_2 r_c \,  \times  
\left( 1+\frac{r_x^2}{\zeta_1^2} + \frac{r_z^2}{\zeta_3^2} 
\right)^{\frac{1}{2}} \nonumber \\ 
& \times & \frac{\left[ B\left(\frac{3}{2}\beta-\frac{1}{2},
\frac{1}{2}\right) -B_m\left(\frac{3}{2}\beta-
\frac{1}{2},
\frac{1}{2}\right)\right]^2}{\left[ B\left(3\beta-
\frac{1}{2},\frac{1}{2}\right) -B_m\left(3\beta-\frac{1}{2},
\frac{1}{2}\right)\right]}~, 
\label{eq:hubble}
\end{eqnarray}

\begin{equation}
{\rm where} \ \ 
\lambda = \frac{4 \pi (1+z)^3}{\epsilon} \times \left[
\frac{k_B \sigma_T}{m_e c^2} \right]^2 ~. 
\nonumber
\end{equation}

%--------------------------------------------------------------------------
\section{Errors obtained in the quantities $\vec{y},~\vec{S_X}$ and $\vec{H_0}$}
\label{SZ:S3}
%--------------------------------------------------------------------------
%-------------------------------------------
\subsection{Finite extension of clusters}
%-------------------------------------------
Since the hot gas in a real cluster has a finite extension, each 
of the observed quantities of the Compton parameter and X-ray surface brightness
will be smaller than that estimated based on the assumption that $l \rightarrow 
\infty$. The incomplete beta-function $B_m$ in Eqs. (\ref{eq:dt}) and 
(\ref{eq:x}) 
can be seen as a correction term due to the finite extension, which together 
with the geometry of the cluster, enters through the $m$-parameter (see Appendices). 
As an illustration we report here the analysis of the influence of this 
correction 
for the simplest cluster case: isothermal $\beta=2/3$-model with a 
spherical density profile (i.e. $\zeta_1=\zeta_2=\zeta_3=1$), and a line of 
sight going through the cluster center (i.e. $r_x=r_z=0$). The reason 
for this choice is to be able to neglect, in this case, 
geometrical effects and thus to focus only on the modification due to the 
finite extension.

\begin{figure}[h]
\begin{center}
\epsfig{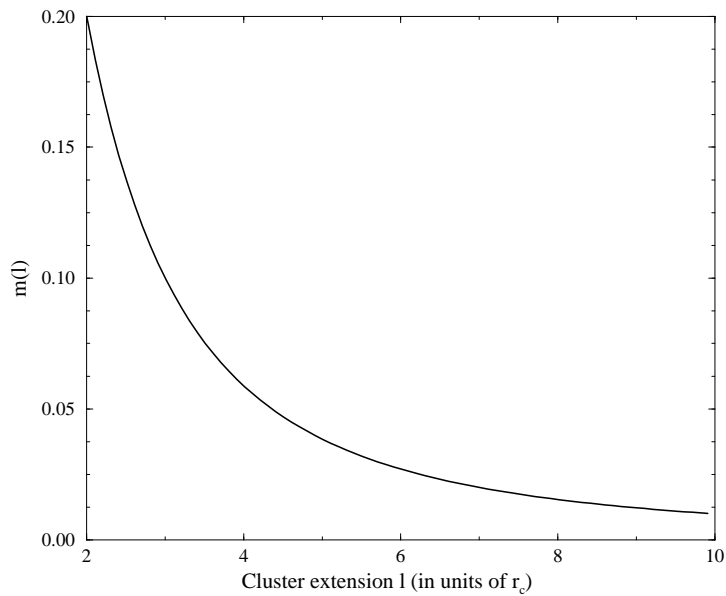}
\caption{Relation between the $m$-parameter of the incomplete beta-function
$B_m$ and the cluster extension $l$. The line of sight goes through
the center of a spherical cluster (isothermal $\beta=2/3$-model).}
\label{fig:ml}
\end{center}
\end{figure}

In Fig. (\ref{fig:ml}) we have plotted the parameter $m$ as a function of 
$l$ for an isothermal $\beta=2/3$-model. 
We notice that high values of $l$ correspond to 
small values of $m$, which is almost vanishing for $l \, > \, 6$, 
and, therefore, the correction for the finite extension becomes negligible 
($m \, \rightarrow \, 0$ for  
$l \, \rightarrow \, \infty$). 

\begin{figure}[h]
\begin{center}
\epsfig{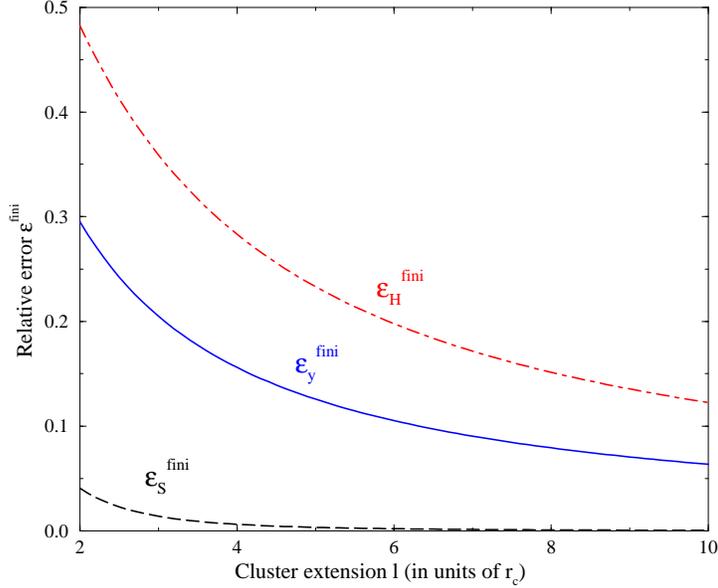}
\caption{Influence of the finite extension on the Compton parameter, 
the surface brightness and the Hubble constant assuming $\beta=2/3$, 
$r_x=r_z=0$ and a spherical cluster. 
The relative errors which are shown are defined in the text.}
\label{fig:extl}
\end{center}
\end{figure}

We denote $y(\infty)$ and $S_X(\infty)$, respectively, as 
the expressions of the Compton parameter and of the X-ray 
surface brightness for a cluster 
with infinite extension, and $y(l)$ and $S_X(l)$ as 
the same expressions for a cluster 
with a finite extension $l$ along the line of sight. The relative errors 
$\epsilon_y^{fini}$ on the Compton parameter and 
$\epsilon_{S}^{fini}$ on the surface brightness 
are defined by the expressions:
\begin{equation}
\epsilon_{y}^{fini} = \frac{y(\infty) - 
y(l)} {y (\infty)} \ \ {\rm and} \ \ 
\epsilon_{S}^{fini}= \frac{S_X (\infty) - S_X (l)}
{S_X (\infty)} ~.
\end{equation}
>From Eqs. (\ref{eq:dt}) and (\ref{eq:x}) we can easily estimate these 
ratios which are
\begin{equation}
\epsilon_{y}^{fini} =
\frac{B_m(3/2\beta-1/2,1/2)}{B(3/2\beta-1/2,1/2)}
\ \ {\rm and} \ \ 
\epsilon_{S}^{fini} =
\frac{B_m(3\beta-1/2,1/2)}{B(3\beta-1/2,1/2)}~.
\end{equation}
For a spherical density profile, a line of sight through the cluster
center and an infinite extension, 
we obtain from Eq. (\ref{eq:hubble}):
\begin{eqnarray}
\label{eq:constant}
\frac{y^2(\infty)}{S_X(\infty)} = 
\lambda T_{eo}^{3/2} \, r_c \, \frac{\left[B(\frac{3}{2}\beta 
-\frac{1}{2}, \frac{1}{2} ) \right]^2}
{ B(3 \beta -\frac{1}{2}, \frac{1}{2} )}~.
\end{eqnarray}
We introduce $\theta_c=r_c/D_A$ the angular core radius, where $D_A$ is 
the angular diameter distance of the cluster:
\begin{equation}
D_A =  \frac{{\cal C}}{H_0} \ \ {\rm where} \ \ 
{\cal C} = \frac{c}{q_0^2} \frac{q_0 z + (q_0 -1)
(\sqrt{1 + 2q_0 z}-1)}{(1+z)^2} .
\end{equation}
With Eq. (\ref{eq:constant}) we can estimate the Hubble constant 
$H_0(\infty)$ for an infinitely extended cluster by: 
\begin{equation}
H_0 (\infty) =  \lambda' \, T_{eo}^{3/2} \, \frac{S_X(\infty)}
{y^2(\infty)} \, \theta_c \, 
\frac{\left[B(\frac{3}{2}\beta 
-\frac{1}{2}, \frac{1}{2} ) \right]^2}
{ B(3 \beta -\frac{1}{2}, \frac{1}{2} )} \ \
{\rm with} \ \ \lambda'=\lambda {\cal C}
\end{equation}
and for a finite extension $l$, we get instead
\begin{equation}
H_0(l) =  
 \lambda' T_{eo}^{3/2} \, \frac{S_X(l)}
{y^2(l)} \, \theta_c \,  
\frac{\left[ B\left(\frac{3}{2}\beta-\frac{1}{2},
\frac{1}{2}\right) -B_m\left(\frac{3}{2}\beta-
\frac{1}{2},
\frac{1}{2}\right)\right]^2}{\left[ B\left(3\beta-
\frac{1}{2},\frac{1}{2}\right) -B_m\left(3\beta-\frac{1}{2},
\frac{1}{2}\right)\right]}.
\end{equation}
Obviously, $S_X$ and $y^2$ are observed quantities and thus the ratios 
$S_X(\infty)/y^2(\infty)$ and $S_X(l)/y^2(l)$ are in the following both set 
equal to the measured value $\Bigl( S_X/y^2 \Bigr)_{obs}$. This way, for our 
example, we get for the relative error for the estimation of the Hubble 
constant, due to assuming an infinite extension rather than a finite one, 
the expression:
\begin{equation}
\epsilon_{H_0}^{fini} = 
\frac{H_0(\infty) - H_0(l)}{H_0(\infty)}
 =  1- 
\frac{ 
B(3 \beta-\frac{1}{2},\frac{1}{2})
\left[
B(\frac{3}{2}\beta-\frac{1}{2},\frac{1}{2})-
B_m(\frac{3}{2}\beta-\frac{1}{2},\frac{1}{2}) 
\right]^2
}
{
B^2(\frac{3}{2}\beta-\frac{1}{2},\frac{1}{2})
\left[
B(3 \beta-\frac{1}{2},\frac{1}{2}) - 
B_m(3 \beta-\frac{1}{2},\frac{1}{2})
\right]
}~.
\label{eq:epsH}
\end{equation}

In Fig. (\ref{fig:extl})  we plot the relative error
as a function 
of the finite extension $l$ for a spherical isothermal $\beta=2/3$-model. 
For $l \rightarrow \infty$ the relative errors on the Compton parameter, 
the X-ray surface brightness and the Hubble parameter become, of course, 
negligible. For instance, for a
cluster with a finite extension of about 10 times $r_c$, 
the relative error with respect to the assumption of an infinite extension
is only about 6 \% for the Compton parameter. 
For the X-ray surface brightness, the relative error due 
to the finite extension is much smaller, for instance an 
error of about 4\% is obtained if the cluster has an
extension of only about 2 times $r_c$. Nevertheless, 
for clusters with a {\it small} extension (3$r_c$) 
with respect to their core radius, 
the error in $\epsilon_y^{fini}$ becomes quite substantial ($\sim 20$\%)
for the Compton parameter.
\\
These estimations are in accordance with the results of Inagaki 
et al. (\cite{Ina95}). The net effect when one considers 
{\it infinite clusters} is thus to {\it overestimate} the value of the 
temperature decrement and the X-ray surface brightness. 
\\
The influence of the finite extension on the Hubble constant given in
Eq. (\ref{eq:epsH}) is larger. Indeed, the Hubble parameter is 
overestimated by almost 20\%, when  considering for instance a cluster with 
infinite extension as compared to one with an extension of 7$r_c$ 
(cf. Fig. \ref{fig:extl}).
%----------------------------------
\subsection{Polytropic index}
\begin{figure}[h]
\begin{center}
\epsfig{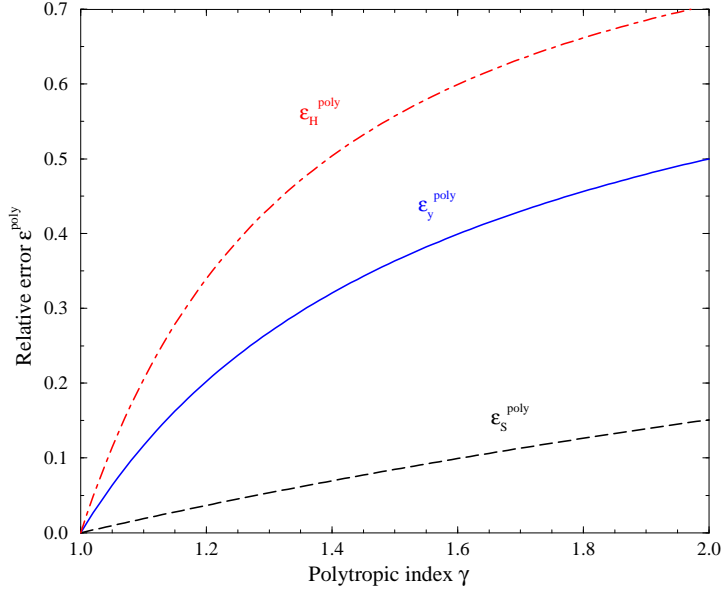}
\caption{Relative error $\epsilon_y^{poly}$ (for 
$\Delta T_{SZ}$), $\epsilon_S^{poly}$ (for $S_{X}$) and 
 $\epsilon_H^{poly}$ for the Hubble constant between 
a polytropic and an isothermal profile. The line of sight 
is taken to go through the center of the cluster, which is 
assumed to have a spherical profile with infinite extension 
($\beta=2/3$-model).}
\label{fig:pgamma}
\end{center}
\end{figure}

%-----------------------------------
Although the isothermal distribution is often a reasonable approximation of 
the actual observed clusters, some clusters do show non-isothermal features. 
Henriksen \& Mushotzky 
(\cite{hen85}) have suggested that the isothermal model cannot be consistently 
applied to gas distributions in clusters. 
Indeed, Markevitch et al. (\cite{Mar98}) find that the 
temperature profiles of some clusters can be 
approximately described by a polytrope. More recently, Ettori et al. 
(\cite{Ett00}), with a combined analysis of the BeppoSAX and ROSAT-PSPC 
observations, showed that a polytropic profile with 
index $\gamma=1.16 \pm 0.03$ fits the temperature 
distribution of the cluster A3562 very closely.\\
As a consequence, within the virial regions of typical clusters of galaxies,
the gravitating mass, the gas mass and the gas fraction can vary quite
substantially, compared to that obtained assuming an isothermal profile 
(Ettori \& Fabian \cite{Ett99}, Ettori \cite{Ett00}, 
Ettori et al. \cite{Eta00}). It is interesting, therefore, to investigate 
the variations due to a polytropic equation of state on SZ quantities.
\\
A polytropic thermal profile has the following form:
\begin{equation}
T_e = T_{eo} \, \left[ \frac{n_{e}}{n_{eo}}\right]^{\gamma-1}~, 
\end{equation}
where the subscript $o$ denotes the values in the cluster center and 
$\gamma$ is the polytropic index. The isothermal profile is obtained by
setting $\gamma=1$.
\\
Then the Compton parameter and surface brightness are given by
\begin{equation}
y^{poly} = 2 \, 
\frac{k_B T_{eo}n_{eo}r_c \sigma_T}{m_e c^2}   
\, \int_{0}^{l} \, 
\left[ \frac{n_e}{n_{eo}} \right]^{\gamma} dr_y 
\end{equation}
\begin{equation}
S_X^{poly}= \frac{\epsilon \sqrt{T_{eo}} n_{eo}^2 r_c}{2 \pi (1 + z)^3}
\, \int_{0}^{l} \, 
\left( \frac{n_e}{n_{eo}} \right)^{\frac{\gamma}{2}+\frac{3}{2}} \, dr_y. 
\end{equation}
Similarly to the calculations in Section \ref{SZ:S2}, we find for $y^{poly}$ 
and $S_X^{poly}$:
\begin{eqnarray}
y^{poly} (r_x,r_z) =  
 & & \frac{k_BT_{eo}  \zeta_2 r_c\sigma_T n_{eo}}
{m_e c^2} \, \times \,
\left( 1+\frac{r_x^2}{\zeta_1^2} + \frac{r_z^2}{\zeta_3^2} 
\right)^{-\frac{3}{2} \beta \gamma + \frac{1}{2}} 
 \nonumber \\
&\times& \,  
\left[ B\left(\frac{3}{2}\beta \gamma -\frac{1}{2},\frac{1}{2}\right) -
B_m\left(\frac{3}{2}\beta \gamma -
\frac{1}{2}, \frac{1}{2}\right)\right]~,
\end{eqnarray}
\begin{eqnarray}
S_X^{poly} (r_x,r_z) = & & 
\frac{\epsilon  \sqrt{T_{eo}}n_{eo}^2\zeta_2 r_c}{4 \pi 
(1 + z)^3} \, \times  \, 
\left( 1+\frac{r_x^2}{\zeta_1^2} + \frac{r_z^2}{\zeta_3^2} 
\right)^{-\frac{3}{2} \beta (\frac{\gamma}{2} +\frac{3}{2}) + \frac{1}{2}}
 \nonumber \\
&\times& \,  \left[ B\left(\frac{3}{2}\beta\left(\frac{\gamma}{2}+\frac{3}{2}\right)
-\frac{1}{2},\frac{1}{2}\right) 
-B_m\left(\frac{3}{2}\beta \left(\frac{\gamma}{2}+\frac{3}{2}\right)-\frac{1}{2},
\frac{1}{2}\right)\right]~.
\end{eqnarray}
The relative error by considering a polytropic profile (index $poly$) as
compared to an isothermal profile (index $iso$), can be expressed as follows, 
assuming for simplicity an infinite extension of the cluster,
\begin{eqnarray}
   \epsilon_{y}^{poly}&=&\frac{y_{iso}-y_{poly}}{y_{iso}} \\
&=&1-\left[1+\left(\frac{r_x}{\zeta_1}\right)^2+\left(\frac{r_z}{\zeta_3}\right)^2
\right]
^{-\frac{3}{2}\beta(\gamma-1)}\times\frac{B\left(\frac{3}{2}\beta\gamma
-\frac{1}{2}
,\frac{1}{2}\right)}{B\left(\frac{3}{2}\beta-\frac{1}{2}
,\frac{1}{2}\right)}\nonumber
\end{eqnarray}
for the Compton parameter and
\begin{eqnarray}
   \epsilon_{S}^{poly}&=&\frac{\left(S_{X}\right)_{iso}-\left(S_{X}\right)_{poly}}
{\left(S_{X}\right)_{iso}} \\
&=&1-\left[1+\left(\frac{r_x}{\zeta_1}\right)^2+\left(\frac{r_z}{\zeta_3}\right)^2
\right]
^{-\frac{3}{4}\beta(\gamma-1)}\times\frac{B\left(\frac{3}{2}\beta
(\frac{\gamma}{2}+\frac{3}{2})-\frac{1}{2}
,\frac{1}{2}\right)}{B\left(3\beta-\frac{1}{2}
,\frac{1}{2}\right)}\nonumber
\end{eqnarray}
for the surface brightness.\\
The corresponding error for the Hubble constant turns out to be
\begin{equation}
   \epsilon_{H}^{poly}=\frac{\left(H_0\right)_{iso}-\left(H_0\right)_{poly}}
{\left(H_0\right)_{iso}} 
=1-\frac{\left[B\left(\frac{3}{2}\beta \gamma
-\frac{1}{2},\frac{1}{2}\right)\right]^2
B(3\beta-\frac{1}{2},\frac{1}{2})}{\left[B\left(\frac{3}{2}\beta-\frac{1}{2},\frac{1}{2}\right)\right]^2
B(\frac{3}{2}\beta(\frac{\gamma}{2}+\frac{3}{2})-\frac{1}{2},\frac{1}{2})}~,
\end{equation}
when taking measurements along the line of sight going through the cluster 
center. The relative errors defined above for polytropic indices 
between 1 and 2 are shown in 
Fig. (\ref{fig:pgamma}).
Already, a small deviation from the isothermal case ($\gamma=1$) leads to
significant relative errors and thus to quite different values of the
observable quantities. For instance, assuming an isothermal 
instead of a polytropic profile with index $\gamma=1.2$ leads 
to an overestimation of 20\%, 4\% and 34\% for the quantities $y,~S_X$ 
and $H_0$, respectively.

\begin{figure}[h]
\begin{center}
\epsfig{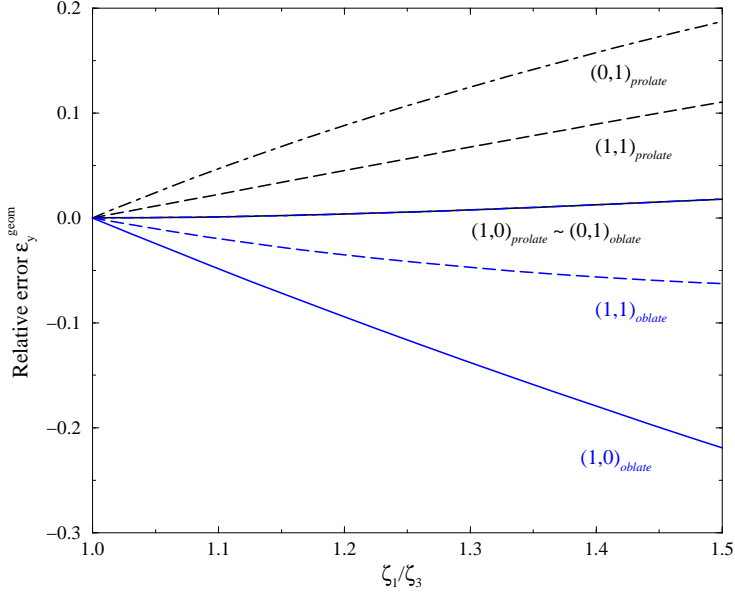}
\caption{Geometrical relative error $\epsilon_y^{geom}$ 
for $y$ between an axisymmetric ellipsoidal and a
spherical geometry (assuming infinite extension, isothermal profile 
and $\beta=2/3$)
for three different lines of sight, parametrized in units of $r_c$. 
The two ellipsoidal
shapes $prolate$ and $oblate$ are indicated.}
\label{fig:ts}
\end{center}
\end{figure}

\begin{figure}[h]
\begin{center}
\epsfig{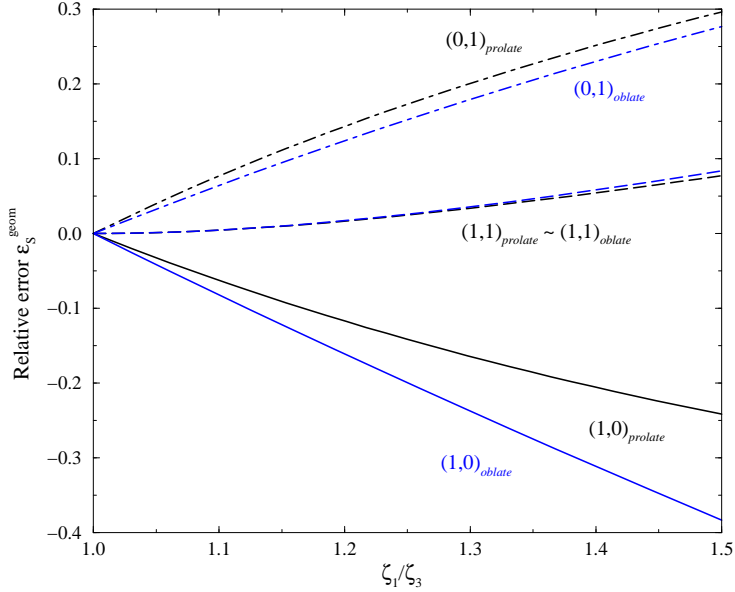}
\caption{Geometrical relative error $\epsilon_{S}^{geom}$ 
for $S_X$ between an axisymmetric ellipsoidal and a spherical geometry 
(assuming infinite extension, isothermal profile and $\beta=2/3$) for three 
different lines of sight, parametrized in units of $r_c$. The two ellipsoidal
shapes $prolate$ and $oblate$ are indicated.}
\label{fig:xs}
\end{center}
\end{figure}

\begin{figure}[h]
\begin{center}
\epsfig{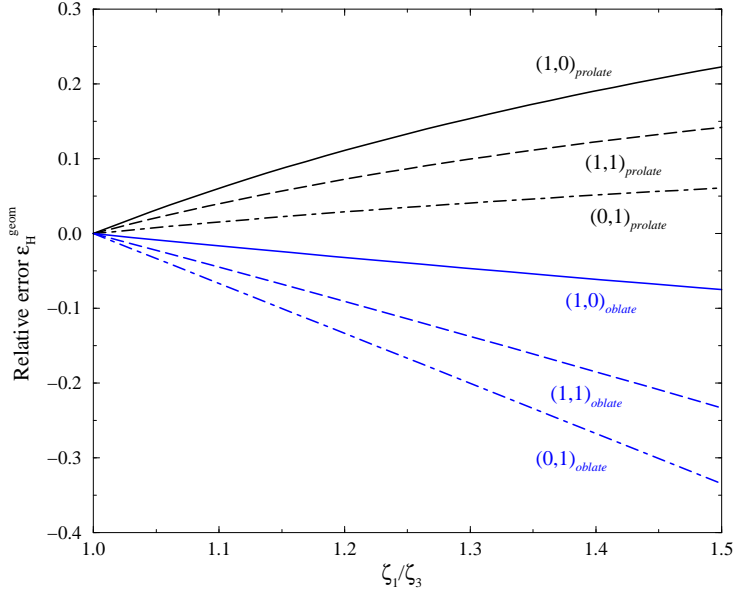}
\caption{Geometrical relative error $\epsilon_H^{geom}$ 
for $H$ between an axisymmetric ellipsoidal and a
spherical geometry (assuming infinite extension, isothermal profile 
and $\beta=2/3$)
for three different lines of sight, parametrized in units of $r_c$. 
The two ellipsoidal
shapes $prolate$ and $oblate$ are indicated.}
\label{fig:hs}
\end{center}
\end{figure}
%------------------------------------
\subsection{Geometrical effect}
%------------------------------------
In order to compare the relative errors induced by geometrical effects, i.e. 
the difference between 
ellipsoidal and spherical geometries, we choose the parameters of the 
cluster for both geometries such that we get the same value for the emission 
integral 
EI, defined by Sarazin (\cite{Sar88}) as:
\begin{equation} 
{\rm EI} =  \int \, n_e^2 \, dV, 
\end{equation}
where $V$ is the volume of the cluster. We consider, for simplicity, the case 
where the various integrals are approximated by assuming an infinite 
extension: after some algebra 
(see Appendix A for the change of variables) we get the following expression 
for the emission 
integral for a spherical geometry:
\begin{equation} 
EI_{sph} =  \pi^{3/2} n_{eo}^2 r_c^3 \, 
\frac{\Gamma(3\beta-\frac{3}{2})}{\Gamma(3\beta)}~,
\end{equation}
and for an ellipsoidal geometry:
\begin{equation} 
EI_{ell} =  \pi^{3/2} n_{eo}^2 \zeta_1 \zeta_2 \zeta_3 r_c^3  \, 
\frac{\Gamma(3\beta-\frac{3}{2})}{\Gamma(3\beta)}~.
\end{equation}
Thus the {\it equal emission} condition $EI_{sph}  =  EI_{ell}$ 
requires $\zeta_1\; \zeta_2\; \zeta_3=1$.\\
In the following we will consider the two axisymmetric cases
{\em prolate} (cigar shaped) with symmetry axis $r_x$, thus 
$\zeta_2=\zeta_3=\sqrt{1/\zeta_1}$, and
{\em oblate} (pancake shaped), where the symmetry axis is given by $r_z$,
and thus $\zeta_2=\zeta_1$ and 
$\zeta_3=1/\zeta_1^2$.\\

We introduce $\epsilon_{y}^{geom}$ and $\epsilon_{S}^{geom}$ 
as the geometrical relative error between ellipsoidal and 
spherical geometry for the Compton parameter and the surface brightness, 
respectively,  
\begin{equation}
\epsilon _{y}^{geom} = \frac{y_{sph} - y_{ell} } { y_{sph}} 
\ \ {\rm and} \ \ 
\epsilon _{S}^{geom} = \frac{ \left( S_{X} \right)_{sph} - 
 \left( S_{X} \right)_{ell} } { \left( S_{X} \right)_{sph}}~, 
\end{equation}
and thus find\footnote{Remember that the set of coordinates ($r_x,r_y,r_z$), 
as well as the half
axes of the ellipsoid ($\zeta_1,\zeta_2,\zeta_3$), are given in units of the 
core radius $r_c$.}
\begin{equation}
\epsilon_{y}^{geom} \, = \,
1- \zeta_2~\left(
\frac{1+\frac{r_x^2}{\zeta_1^2}+\frac{r_z^2}{\zeta_3^2}}
{1+r_x^2 + r_z^2}
\right)^{-\frac{3}{2}\beta+\frac{1}{2}} \, \, ~{\rm and}
\end{equation}
\begin{equation}
\epsilon_{S}^{geom} = 
1-\zeta_2~\left(
\frac{1+\frac{r_x^2}{\zeta_1^2}+\frac{r_z^2}{\zeta_3^2}}
{1+r_x^2+ r_z^2}
\right)^{-3\beta+\frac{1}{2}}~.  
\end{equation}
In Figs. (\ref{fig:ts}) and (\ref{fig:xs}) we have plotted 
the 
geometrical relative error as a function of the flattening of the profile 
$\zeta_1/\zeta_3$ for the Compton 
parameter $\epsilon_{y}^{geom}$ and the surface brightness  
$\epsilon_{S}^{geom}$, respectively.
We consider three different lines of sight, $(r_x,r_z )$ = (1,0), (1,1) and 
(0,1), given in units of the core radius $r_c$.\\
Fig. (\ref{fig:ts}) shows, that the line of sight (1,0) in the prolate case
leads to almost the same error of the Compton parameter as the
line of sight (0,1) in the oblate case. For a flattening of 50\% (i.e.
$\zeta_1/\zeta_3~=~1.5$) we get an overestimation  of about 2\% by using 
a spherical cluster shape instead of an ellipsoidal one. The other 
lines of sight lead in the prolate case to overestimations of up to 19\%, while
in the oblate case the Compton parameter is underestimated by almost 22\%.\\
Due to a quadratic dependance on the density profile, the surface brightness 
is more affected by a flattened shape. The line of sight (1,1) shows in both
cases (prolate and oblate) an overestimation of about 8\%. Towards (0,1) the
surface brightness is overestimated in both cases by up to 30\%, while the
view towards (1,0) results in an underestimation of 24\% for a prolate 
deformation and  almost 38\% in the oblate case.

In the case of an infinite extension approximation from Eq. (\ref{eq:hubble}) 
we get 
\begin{equation}
\frac{y^2(r_x, r_z)}{S_X (r_x,r_z)} =  
\lambda \, T_{eo}^{3/2} \, \zeta_2 r_c \, \times  
\left( 1+\frac{r_x^2}{\zeta_1^2} + \frac{r_z^2}{\zeta_3^2} 
\right)^{\frac{1}{2}} \times  
\frac{B^2\left(\frac{3}{2}\beta-\frac{1}{2},
\frac{1}{2}\right)}{B\left(3\beta-\frac{1}{2},\frac{1}{2}\right)}~.
\end{equation}
Again, the ratio $y^2/S_X$ ratio is set equal to the observed value and
 leads then to the relative error on the Hubble constant:
\begin{equation}
\epsilon_{H}^{geom} = \frac{ \left( H_0 \right)_{sph} - 
 \left( H_0 \right)_{ell} } { \left( H_0 \right)_{sph}} =  
1-\zeta_2~\left(
\frac{1+\frac{r_x^2}{\zeta_1^2}+\frac{r_z^2}{\zeta_3^2}}
{1+r_x^2+ r_z^2}
\right)^{\frac{1}{2}}~.
\end{equation}
Fig. (\ref{fig:hs}) shows in the axisymmetric ellipsoidal cases, prolate 
and oblate, the influence of a flattening of the cluster profile up to 
50\%. The prolate case results in a systematic overestimation of the Hubble
constant, by considering a spherical instead of a flattened profile. Depending
on the line of sight, this error goes up to 22\%. Underestimations of up to
33\% arise in the oblate case.\\
We summarize in Table (\ref{tab:geom}) the relative errors
on the $y$-parameter, the surface brightness and the Hubble constant 
that appear by considering a spherical instead of a flattened ellipsoidal
shape with, as an example, an axis ratio of 1.2. Negative values indicate that
the considered quantity is underestimated, whereas positive values that it
is overestimated.

\begin{table*}[h!tbp]
 \renewcommand{\arraystretch}{1.0}
  \centering
    \begin{tabular}{|l|c||c|c|c|}\hline
    Cluster shape    &  Line of sight  &  $\epsilon _{y}^{geom}$& 
                     $\epsilon _{S}^{geom}$&$\epsilon _{H}^{geom}$ \\ 
                     &   (in $r_c$)    & (in \%)                 &
                      (in \%)                 &(in \% )     \\ \hline\hline 
 {\em prolate}   &(1,0)&  0.4    &-11.7&11.1 \\ \hline
   &(1,1)&4.5 &1.7 &7.3  \\ \hline
   &(0,1)&8.8 & 14.3& 2.9 \\ \hline
 {\em oblate}   &(1,0)& -9.4     &-16.1& -3.2\\ \hline
   &(1,1)&-3.5 & 1.7&-9.1  \\ \hline
   &(0,1)&0.4 & 12.4& -13.3 \\ \hline

    \end{tabular}
    \caption{\small{Relative errors on the $y$-parameter, the surface 
brightness and the Hubble constant are shown. The flattened ellipsoidal is 
supposed to have an axes ratio of 1.2. Negative numbers indicate 
underestimations, whereas positive ones overestimations.}}
    \label{tab:geom}
\end{table*}

In the general triaxial case (i.e. not necessarily prolate or oblate), 
estimating the Hubble constant is more difficult.  
We can compute the angular diameter distance $D_A$ and thus
the Hubble constant by measuring the angles $\theta_1$ and 
$\theta_3$, defined as the angular ellipsoidal core radii 
$\theta_1=\zeta_1 r_c/D_A$ and $\theta_3=\zeta_3 r_c/D_A$, but we have no
observational access to the corresponding angle $\theta_2$. 
On the contrary, in the case of a 
spherical profile it is sufficient to measure, as we have seen in section 
(3.1), the value of $y$ and $S_X$ towards the center of the 
cluster in order to evaluate the Hubble constant.
We do not discuss further the geometrical relative error for a general 
triaxial cluster. 
%-------------------------------------
\subsection{Projection effect}
%-------------------------------------
Ruiz (\cite{Rui76}) and Stark (\cite{Sta77}) have 
discussed the projection onto the sky of 
luminosity distributions which have an 
ellipsoidal form. This problem has also been treated by Fabricant, 
Rybicki \& Gorenstein (\cite{Fab84}). The projection effect is expected to broaden the 
measurements of the temperature 
decrement and the surface brightness as mentioned by Cooray 
(\cite{Coo98}). 
\\
In order to evaluate the projection effect we assume an 
ellipsoidal geometry. As an
example, we take the case of an infinite cluster extension with 
isothermal profile and a prolate form.
\\
The profile of the cluster is supposed to be rotated by an angle 
$\theta$ around the $r_z$ axis. For comparison, we consider a prolate profile
with a major half axis along $r_x$, which has the same projected image on the 
sky, i.e. on the ($r_x,r_z$)-plane. 
We then compute the difference on the resulting $y$-parameter and surface 
brightness between these two profiles. The structure
integrals $I^{proj}$ and $I^{rot}$ are given in Appendix B, where 
we have computed the $y$ parameter and the surface brightness of a cluster 
both for a rotated coordinate system, with an angle $\theta$, 
and by considering its projection onto the sky.
\\ 
The relative error can be quantified as
\begin{equation}
\epsilon _{I}^{proj}  =  \frac{I^{proj}-I^{rot}}{I^{proj}}~.
\end{equation}
It is a pure geometrical effect and thus the same 
for the $y$ parameter, the surface brightness and the Hubble constant.\\

\begin{figure}[h]
\begin{center}
\epsfig{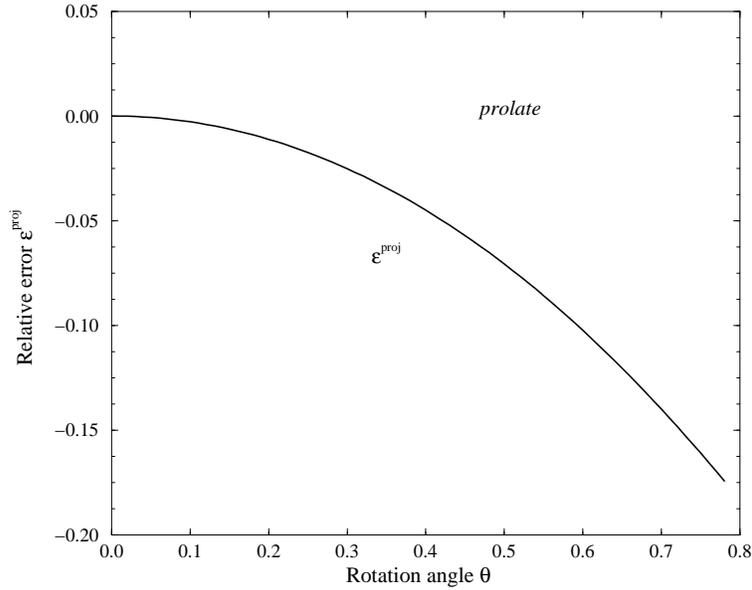}
\caption{The relative error, when comparing a projected and a rotated 
prolate-shaped cluster, is shown as a 
function of $\theta$. The axis ratio is fixed to be $\zeta_1/\zeta_3=1.5$.}
\label{fig:deptheta}
\end{center}
\end{figure}

\begin{figure}[h]
\begin{center}
\epsfig{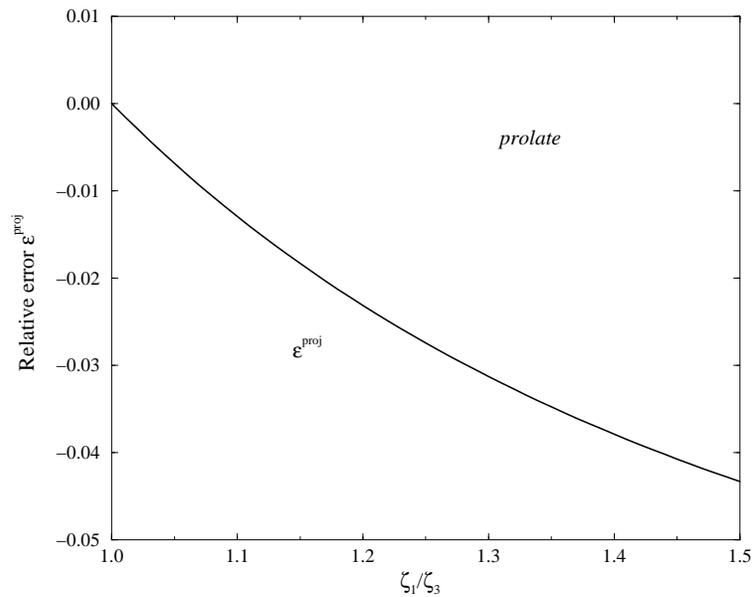}
\caption{The relative error between a projected and a rotated prolate-shaped
cluster profile is shown as a 
function of the axes ratio $\zeta_1/\zeta_3$. The rotation angle $\theta$ is 
fixed at $\pi/4$.}
\label{fig:deps1}
\end{center}
\end{figure}

In Figs. (\ref{fig:deptheta}) and (\ref{fig:deps1}) we have plotted the 
relative error due to the projection effects as a 
function of the rotation angle $\theta$ and the axes ratio $\zeta_1/\zeta_3$
of the ellipsoid.

\noindent
The maximal $\theta$ is assumed to be $\pi/4\sim 0.785$, for which we find 
an underestimation of almost
17\%. The influence of the axes ratio turns out to be less than 5\%. 

%%%%%%%%%%%%%%%%%%%%%%%%%%%%%%%%%%%%%%%%%%%%%%%%%%%%%%%%%%%
\section{Discussion and conclusions}
\label{SZ:S4}
%%%%%%%%%%%%%%%%%%%%%%%%%%%%%%%%%%%%%%%%%%%%%%%%%%%%%%%%%%%%

We have calculated the relative error caused by assumptions regarding 
finite extension, a polytropic 
temperature profile, ellipsoidal geometry and projection effects, on the 
measurements of the X-ray surface brightness, the SZ temperature decrement
and the determination of the Hubble constant. 
Although the X-ray data have improved dramatically in the last decade, it 
is still difficult to determine the internal structure of clusters from 
X-ray imaging alone, because such images supply only 
projected temperature and surface 
brightness information, without further indications of the 
internal gas dynamics. Nevertheless, recent observations show indirectly 
that 
many clusters are still dynamically evolving (see Mohr 
et al. \cite{mo95}).\\
Cooray (\cite{Coo2000}) has discussed 
intrinsic cluster shape, in particular considering 
axisymmetric models such as oblate and prolate ellipsoids, using the Mohr 
et al. (\cite{mo95}) 
cluster sample. Their study shows that clusters do indeed have 
aspherical profiles, which are more likely described as 
prolate rather than oblate
ellipsoids. Nevertheless, Mohr et al. (\cite{mo95}) remarked 
that they cannot rule out 
the possibility
that clusters are intrinsically triaxial.

Pierre et al. (\cite{Pie96}) studied with ROSAT 
the rich lensing cluster Abell 2390 and determined its gas and matter 
content. They found that on large scales the 
$X$-ray distribution has an elliptical shape with an axis ratio of 
minor to major half 
axis of $\zeta_1/\zeta_3\sim 1.33$. Using our results we see that this 
corresponds to a 
relative error in the $y$ parameter of up to 10\%, depending on the line of 
sight
and the shape of the cluster (prolate or oblate, see Fig. \ref{fig:ts}). The 
surface brightness measurements lead to errors of up to
25\% (see Fig. \ref{fig:xs}) and thus the Hubble constant is 
overestimated by about 23\% (see Fig. \ref{fig:hs}).

An unresolved temperature gradient in the gas affects the gas profile and
thus the total mass derived assuming hydrostatic equilibrium. 
If such a gradient is present, the true 
temperature in the central region may be higher than the emission-weighted 
temperature generally used. As an example, Grego et al. 
(\cite{gre2000}) observed in Abell 370 
a slow decline of the temperature with radius. The temperature 
falls to half its central value within 6-10 core radii. This 
temperature profile can be approximately described by a gas with a 
polytropic index of $\gamma=1.2$, 
which in itself is already an important modification with respect to an isothermal 
profile and could lead to a 
relative error 
$\epsilon^{poly}_H$ of 37\%  in the evaluation of the Hubble constant 
(see Fig. \ref{fig:pgamma}).
\\
Furthermore, the optical and X-ray observations 
of this cluster show a possible bimodal mass distribution. Thus, the combined 
temperature and geometry effects must be taken into account to obtain 
reliable values for such parameters as the gas and total matter content. 
A similar 
polytropic index ($\gamma =1.16$) has also been found for Abell 3562 
(Ettori et al. \cite{Ett00}).

\begin{figure}[h]
\begin{center}
\epsfig{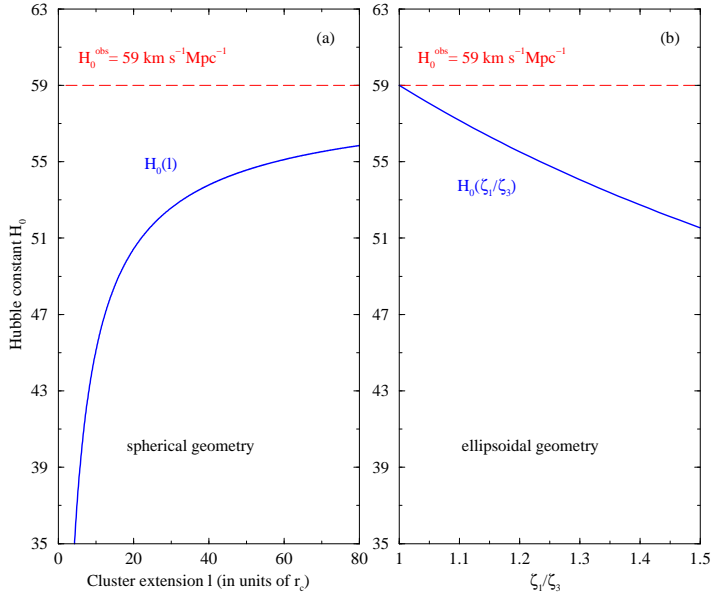}
\caption{The Hubble constant derived from the data of Mauskopf 
et al. (2000). Fig. (9a) shows the influence of finite extension, 
while Fig. (9b) gives the value of $H_o$ assuming an axisymmetric ellipsoidal 
geometry. In the latter case, oblate or prolate geometry give the same 
value of $H_o$ when taking a line of sight through the cluster center, 
as is assumed here.}
\label{fig:obs1}
\end{center}
\end{figure}

Cooling flows in galaxy clusters can substantially change the temperature 
profiles, especially in the inner regions. 
Schlickeiser (\cite{Sch91}) and Majumdar \& Nath (\cite{maju2000}) 
have investigated the changes induced by a cooling flow in the temperature 
and density profiles, and their implications on the SZ effect. We notice that 
for a polytropic distribution, the density profile can still be well 
approximated by a $\beta$ profile, whereas for cooling flow solutions 
the density becomes quite different. For example, Vikhlinin et al. 
(\cite{vik99}) showed that outside the cooling flow region, the 
$\beta$-model describes the observed surface brightness closely, but not 
precisely. In this context,  Majumdar \& Nath (\cite{maju2000}) found that the 
presence of a cooling flow in a cluster can lead to an overestimation of the 
Hubble constant determined from the SZ decrement.
\vskip2mm
Recently, Mauskopf, Ade, Allen et al. (\cite{maus2000}) determined the Hubble 
constant from $X$-ray measurements obtained of the cluster Abell 1835 
with ROSAT and from the corresponding millimetric observations of the 
SZ effect with 
the Sunyaev-Zel'dovich Infrared Experiment (Suzie) multifrequency array 
receiver. Assuming an infinitely extended, spherical gas distribution with an 
isothermal equation of state, characterized by $\beta=0.58 \pm 0.02$, 
$T_{eo}=9.8 ^{+2.3}_{-1.3}$ keV and $n_{eo} = 5.64 ^{+1.61}_{-1.02} 
\times 10^{-2}$ cm$^{-3}$, they found a value of 
$H_o=59^{+38}_{-28}$ km s$^{-1}$ Mpc$^{-1}$ for the Hubble constant. 
In Fig. (\ref{fig:obs1}) we 
show the influence of geometry and of assumptions of finite extension on this result 
using the same input parameters. 
Fig. (9a) shows that for a spherical geometry, $H_o$ displays a strong dependence on the 
cluster extension. Fig. (9b) gives the value of $H_o$ assuming 
an infinite extended ellipsoid shaped cluster (instead of a spherical 
geometry), as a function of its axis ratio $\zeta_1/\zeta_3$.

\vskip2mm
\noindent
In summary, we see that it is crucial to know the shape of a cluster and 
its temperature profile. For this problem, the new X-ray satellites have
the necessary spatial and spectral resolution to remove the effects of 
contaminating sources in the field and to measure 
the spatial variation of the cluster temperature. In this context it will be 
better for future studies to focus on nearby cluster samples, which are less 
subject to observational selection effects (as mentioned by Roettiger et al., 
(\cite{Roe97})).

\begin{acknowledgements}
We thank Andreas Obrist, Sabine Schindler and Yoel Rephaeli for 
interesting discussions. We are grateful to the referee for useful 
comments. This work has been supported 
by the {\it D$^r$ Tomalla Foundation} and by the Swiss National Science 
Foundation. 
\end{acknowledgements}
\clearpage
\appendix
\section{}
{\bf Structure integral for the $\vec{y}$ parameter}
\\
The structure integral for the SZ effect is given by the expression
\begin{equation}
I_{SZ}(r_x,r_z) = \, 2 \, \int_{0}^{l} \, \frac{n_e}{n_{eo}} dr_y,   
\end{equation}
where $n_e$ is the density profile of the electrons, $n_{eo}$ the central 
density of the cluster and $l$ the maximum extension of the hot gas on the 
line of sight $r_y$, in units of the core radius $r_c$. 
With an ellipsoidal density profile, we obtain
\begin{equation}
I_{SZ}(r_x, r_z) \, = \, 2 \, \int_{0}^{l} \, \left[ 1+ 
\frac{r_x^2}{\zeta_1^2} + \frac{r_y^2}{\zeta_2^2} + \frac{r_z^2}{\zeta_3^2}
\right]^{-\frac{3}{2}\beta} \, dr_y~,
\end{equation}
where $r_i,~\zeta_i$ and $l$ are given in units of the core radius $r_c$.
With the definition of a function $G$ and transforming the variable of 
integration from $r_y$ to $\xi$ such that 
$$
G(\xi) =  \left( 1 + \xi \right)^{-3\beta /2} \ {\rm with}  \ \ 
\xi = \frac{r_x^2}{\zeta_1^2} + \frac{r_y^2}{\zeta_2^2} + \frac{r_z^2}{\zeta_3^2}~,
$$
we obtain after some algebra the structure integral
\begin{equation}
I_{SZ}(r_x,r_z) =  \int_{\eta}^{\eta +(l/\zeta_2)^2} 
G(\xi)\,  ( \xi -\eta )^{-1/2} \, d\xi~, \ \ {\rm where} \ \ 
\eta =  \frac{r_x^2}{\zeta_1^2} + \frac{r_z^2}{\zeta_3^2}~.
\end{equation}
Finally, with a last change of variable,
$$ 
\alpha=\frac{1 + \eta}{1 + \xi}~,
$$
we get
\begin{equation}
I_{SZ}(r_x,r_z) = \zeta_2~(1+ \eta)^{-3(\beta +1)/2} \, 
\int_m ^1 \, \alpha^{(3\beta-3)/2} \, \, (1-\alpha)^{-1/2} \, d\alpha
\end{equation}
with
$$
m= \frac{1 + \eta}{1+\eta + (l/\zeta_2)^2}~.
$$
The structure integral turns out to be
\begin{equation}
I_{SZ} (r_x,r_z) =  \zeta_2~(1+ \eta)^{-\frac{3}{2}\beta+\frac{1}{2}}~
\left[B\left(\frac{3}{2} \beta-\frac{1}{2},\frac{1}{2}\right)-B_m\left(\frac{3}{2} \beta-\frac{1}{2},\frac{1}{2}\right)
\right]~,
\end{equation}
where we introduced the Beta and the incomplete Beta functions defined by
$$
B(a,b)=\frac{\Gamma(a)\Gamma(b)}{\Gamma(a+b)}
$$
and
$$
B_m(a,b)=\frac{\Gamma_m(a)\Gamma_m(b)}{\Gamma_m(a+b)}~,
$$
with the Gamma and the incomplete Gamma-functions $\Gamma$
and $\Gamma_m$, respectively.

\vskip3mm
\noindent
{\bf $\vec{X}$-ray surface brightness}
\\
The structure integral for the $X$-ray surface brightness is given by
\begin{equation}
I_{S_X} =  2 \, \int_{0}^{l} \, 
\left[ \frac{n_e}{n_{eo}}\right]^2 dr_y.  
\end{equation}
With the same transformations, as given above, we get
\begin{equation}
I_{S_X}(r_x,r_z) = \zeta_2~
(1 + \eta)^{-3\beta+\frac{1}{2}} \, 
\left[B\left(3\beta-\frac{1}{2},\frac{1}{2}\right) -   
B_m\left(3\beta-\frac{1}{2},\frac{1}{2}\right) \right]~.
\end{equation}
%---------------------------------------------------------------
\section{}
{\bf Projection effects on the $\vec{y}$ parameter}
%---------------------------------------------------------------
\vskip2mm
\noindent
The rotation in the ($r_x,r_y$)-plane around $r_z$ with an angle $\theta$ 
leads to the density profile:
\begin{equation}
n_e^{rot}(r_x,r_y,r_z,\theta) = n_{eo} \left[ 1+ 
\frac{(r_x \, {\rm cos}\theta +r_y \, {\rm sin}\theta)^2}{\zeta_1^2} + 
\frac{(r_y \, {\rm cos}\theta - r_x \, {\rm sin}\theta )^2}{\zeta_2^2}
+ \frac{r_z^2}{\zeta_3^2} 
\right]^{-3 \beta /2}.
\end{equation}
The rotation angle $\theta$ is the angle between the major half axis of the 
rotated ellipse in the $(r_x,r_y)$ plane and the $r_x$-axis. The line of 
sight is taken to be along the $r_y$-axis.
\\
To investigate the projection effects we will compare a rotated cluster
with respect to the $r_z$ axis with its projection on the ($r_x,r_z$)
plane, corresponding to the sky plane. We assume an infinite extension and 
an isothermal profile. 
\\
The {\em rotated} structure integral for the $y$ parameter turns out to be 
\begin{equation}
I_{SZ}^{rot}(r_x, r_z,\theta ) = \int_{-\infty}^{\infty} \, \left[ 1+ 
\frac{(r_x \, {\rm cos}\theta +r_y \, {\rm sin}\theta)^2}{\zeta_1^2} + 
\frac{(r_y \, {\rm cos}\theta - r_x \, {\rm sin}\theta )^2}{\zeta_2^2}
+ \frac{r_z^2}{\zeta_3^2} 
\right]^{-3\beta /2 } \, dr_y~.
\end{equation}
After some algebra and with the same kind of variable changes as in 
Appendix A, we get (assuming an infinite cluster extension 
for the structure integral):
\begin{equation}
I_{SZ}^{rot}(r_x,r_z,\theta)=\frac{1}{
\sqrt{\frac{{\rm cos}^2\theta}{\zeta_2^2}+\frac{{\rm sin}^2\theta}{\zeta_1^2}}
}~(1+\eta_{\theta})^{-\frac{3}{2}\beta+\frac{1}{2}}~
B\left(\frac{3}{2}\beta-\frac{1}{2},\frac{1}{2}\right)~,
\end{equation}
with
$$
\eta_{\theta}=
\frac{r_x^2}{\zeta_1^2 {\rm cos}^2\theta + \zeta_2^2 {\rm sin}^2\theta}
+\frac{r_z^2}{\zeta_3^2}~.
$$
On the other hand, the projection of this cluster on the observed sky plane 
in the infinite cluster extension case leads to the density profile
\begin{equation}
n_e^{proj}(r_x,r_y,r_z) = n_{eo} \left[ 1+ 
\frac{r_x^2}{\tilde{\zeta}_1^2} + 
\frac{r_y^2}{\tilde{\zeta}_2^2}
+ \frac{r_z^2}{\tilde{\zeta}_3^2} 
\right]^{-3 \beta /2}~,
\end{equation}
where $\tilde{\zeta_1}$ is the maximum value that we get along the $r_x$ axis 
in units of $r_c$
$$
\tilde{\zeta}_1=\sqrt{\zeta_1^2\cos^2\theta+\zeta_2^2\sin^2\theta}~,
$$
moreover, $\tilde{\zeta_3}/\tilde{\zeta_2}=\zeta_3/\zeta_2=1$ (prolate).\\
The {\em projected} structure integral for the $y$ parameter turns out to be 
\begin{equation}
I_{SZ}^{proj}(r_x,r_z,\theta)=\tilde{\zeta_2}~(1+\tilde{\eta_\theta})^
{-\frac{3}{2}\beta+\frac{1}{2}}~
B\left(\frac{3}{2}\beta-\frac{1}{2},\frac{1}{2}\right)
\end{equation}
with
$$
\tilde{\eta_\theta}
=\frac{r_x^2}{\tilde{\zeta_1}^2}+\frac{r_z^2}{\tilde{\zeta_3}^2}~.
$$

\vskip3mm
\noindent
{\bf $\vec{X}$-ray surface brightness}
\vskip2mm
\noindent
For the structure integral of the $X$-ray surface brightness we find
in the case of the {\em rotated} cluster
\begin{equation}
I_{S_X}^{rot}(r_x,r_z,\theta)=\frac{1}{
\sqrt{\frac{{\rm cos}^2\theta}{\zeta_2^2}+\frac{{\rm sin}^2\theta}{\zeta_1^2}}
}~(1+\eta_{\theta})^{-3\beta+\frac{1}{2}}~
B\left(3\beta-\frac{1}{2},\frac{1}{2}\right)~,
\end{equation}
and for the {\em projected} case
\begin{equation}
I_{S_X}^{proj}(r_x,r_z,\theta)=\tilde{\zeta_2}~(1+\tilde{\eta_\theta})^
{-3\beta+\frac{1}{2}}~
B\left(3\beta-\frac{1}{2},\frac{1}{2}\right)~.
\end{equation}
%--------------------------------------------------------------
%  REFERENCES
%--------------------------------------------------------------
{}
\end{document}